\documentclass[draftcls,11pt,onecolumn,oneside]{IEEEtran}
\usepackage{pifont}
\usepackage{times,amsmath,color,amssymb,graphicx,epsfig,cite,psfrag,subfigure}
\usepackage{mathrsfs}
\usepackage{amsfonts}

\begin{document}
\title{Price-Based Resource Allocation for Spectrum-Sharing Femtocell Networks: A Stackelberg Game Approach\vspace{-3mm}}
\author{\authorblockN{Xin Kang$^\dag$, Rui Zhang$^{\dag\ddag}$,  and Mehul Motani$^\dag$}
\authorblockA{\\$^\dag$ Department of Electrical
and Computer Engineering, National University of Singapore,
Singapore 117576\\
Email: \{kangxin, elezhang, elemm\}@nus.edu.sg\\
$^\ddag$ Institute for Infocomm Research, 1 Fusionopolis Way,
$\sharp$21-01 Connexis, South Tower, Singapore 138632\\
Email: rzhang@i2r.a-star.edu.sg}} \maketitle

\begin{abstract}
This paper investigates the \emph{price-based} resource allocation
strategies for the uplink transmission of a spectrum-sharing
femtocell network, in which a central macrocell is underlaid with
distributed femtocells, all operating over the same frequency band
as the macrocell. Assuming that the macrocell base station (MBS)
protects itself by pricing the interference from the femtocell
users, a \emph{Stackelberg game} is formulated to study the joint
utility maximization of the macrocell and the femtocells subject to
a maximum tolerable \emph{interference power constraint} at the MBS.
Especially, two practical femtocell channel models: \emph{sparsely
deployed scenario} for rural areas and \emph{densely deployed
scenario} for urban areas, are investigated. For each scenario, two
pricing schemes: \emph{uniform pricing} and \emph{non-uniform
pricing}, are proposed. Then, the Stackelberg equilibriums for these
proposed games are studied, and an effective \emph{distributed
interference price bargaining} algorithm with guaranteed convergence
is proposed for the uniform-pricing case. Finally, numerical
examples are presented to verify the proposed studies. It is shown
that the proposed algorithms are effective in resource allocation
and macrocell protection requiring minimal network overhead for
spectrum-sharing-based two-tier femtocell networks.
\end{abstract}

\begin{keywords}
Distributed power control, femtocell networks, Stackelberg game,
spectrum sharing, interference management, game theory.
\end{keywords}

\section{Introduction}
As one of the most promising technologies for improving the indoor
experience of cellular mobile users, femtocell has attracted
considerable attentions since it was first proposed. A femtocell is
enabled by a \textbf{home base station} (HBS) that is connected to
the service provider via the third party backhaul (e.g. DSL or cable
moderm). HBSs, also known as Home NodeBs, are short-range low-power
base stations deployed and managed by the customers at home or in
the offices \cite{VChan-Sep2008}. With the help of such HBS,
femtocell users can experience better indoor voice and data
reception, and lower their transmit power for prolonging battery
life. From the network operator's perspective, HBS offsets the
burden on the \textbf{macrocell base station} (MBS), consequently
improving the network coverage and capacity.

In practice, a two-tier femtocell network is usually implemented by
sharing spectrum rather than splitting spectrum between tiers. This
is due to the following reasons: (i) Scarce availability of
spectrum; (ii) Absence of coordination between the macrocell and
femtocells on spectrum allocation; (iii) High requirement on mobile
devices (which need to support switching between different bands in
the splitting-spectrum approach). Therefore, it is more favorable to
operate the macrocell and femtocells in a shared-spectrum from
either an infrastructure or spectrum availability perspective.
However, for spectrum-sharing two-tier femtocell networks, the
cross-tier and inter-cell interference greatly restrict the network
performance. Therefore, the interference mitigation in two-tier
femtocell networks has become an active area of research. A great
deal of scholarly work has recently appeared in the literature on
the design of power control and interference mitigation strategies
for spectrum-sharing femtocell networks. In \cite{ClaussenH}, a
self-configuration transmit power allocation strategy based on the
measured received signal power level from the MBS was developed.  In
\cite{VChan-Aug2009}, the authors proposed a distributed
utility-based Signal-to-Interference-plus-Noise Ratio (SINR)
adaptation algorithm to alleviate the cross-tier interference. In
\cite{HanShinJo-Oct2009},  the authors proposed interference
mitigation strategies in which femtocell users adjust their maximum
transmit power to suppress the cross-tier interference to the
macrocell. In \cite{OFDMA-Femtocell2009}, OFDMA-based femtocell
networks were proposed to manage the interference between macrocell
and femtocells. In \cite{SungsooPark2010}, a macrocell beam subset
selection strategy, which is able to maximize the throughput of the
macrocell, was proposed to reduce the cross-tier interference
between the macrocell and femtocell users. In \cite{Rangan}, to
manage the cross-tier interference and minimize the interference
coordination communication between the macrocell and femtocells, an
effective interference control scheme was proposed to partition the
macrocell's bandwidth into subbands and allow the femtocell users
adaptively allocate power over the subbands. In
\cite{YoungjuKim2010}, the capacity of a two-tier femtocell network
was studied with a practical interference suppression technology. In
\cite{MDohler2010}, a distributed Q-learning algorithm that requires
minimum network overhead and maximizes the network performance was
proposed to manage the interference in femtocell networks.

On the other hand, spectrum sharing with interference control is not
unique to femtocell networks, since it is also an important design
approach for cognitive radio networks (CRNs). In a CRN, secondary
users are allowed to transmit over the frequency bands of primary
users as long as their resulted aggregate interference is kept below
an acceptable level. This constraint is known as \emph{interference
temperature constraint} or \emph{interference power constraint}
\cite{Haykin2005}. With secondary users designing resource
allocation strategies subject to such an interference power
constraint, the interference received at the primary user is
effectively controlled. A great deal of power allocation polices and
interference control strategies have been proposed for
spectrum-sharing CRNs. For example, the optimal power allocation
strategies to maximize the capacity of the secondary user with an
effective protection of the primary user were studied in
\cite{kangTWC, XKang-JSAC} for spectrum-sharing CRNs. The
transmission-capacity tradeoff in a spectrum-sharing CRN was
investigated subject to an outage constraint in \cite{KHuang-JSAC}.
Power and rate control strategies for spectrum-sharing cognitive
radios were studied via dynamic programming under the interference
temperature constraint in \cite{SCui2009}. The spectrum-sharing
problems for CRNs have also been extensively studied via game
theory. In \cite{SKJayaweera-Jul2010}, the authors developed a fair
and self-enforcing dynamic spectrum leasing mechanism via power
control games. Game-theory-based power control strategies to
maximize the utility for spectrum-sharing CRNs were also
investigated in \cite{Bwang-Jul2009} using Stackelberg game, in
\cite{DNiyato-Jul2008} using repeated Cournot game, and in
\cite{EHossain-DNiyato-Aug2009} using evolutionary game,
respectively.

Interference power constraint has been proved to be a practically
useful technique to control the interference in spectrum-sharing
CRNs. However, to the best of the authors' knowledge, it has not
been applied to the design of interference control strategies for
femtocell networks. The main difficulty for such an application lies
in the following fact: Unlike the cognitive radio devices, the
femtocell users are ordinary mobile terminals that may not have the
environment-aware sensing and self power-adaptation capabilities to
control the interference to the macrocell or other underlaid
femtocells. Therefore, imposing interference power constraints at
the femtocell user side to implement the interference control in
femtocell networks becomes unpractical. In this paper, by exploiting
the unique feature of femtocell networks, we apply the interference
power constraint to the design of interference control for the
uplink transmission of femtocell networks in a new way: Instead of
imposing interference power constraints at the femtocell user side,
we assume that such constraints are imposed by the MBS, which
controls the received interference through pricing the interference
from femtocell users. The corresponding \emph{interference prices}
are sent to femtocell users through the existing backhaul links
between the MBS and HBSs. This way, femtocell users are able to
design their power allocation strategies in a decentralized manner
based on the interference prices received from their own HBSs.
Comparing to existing approaches in the literature, our proposed
method perfectly controls the cross-tier interference for femtocell
networks, and at the same time greatly reduces the complexity of
resource allocation implemented by the femtocell users.

The main contributions of this paper are summarized as follows:

\begin{itemize}
\item By bringing the interference power constraint concept from CRNs to the design
of the uplink cross-tier interference control for the two-tier
spectrum-sharing femtocell networks, this paper proposes a new
\emph{price-based} resource allocation scheme for femtocell users,
whereby the MBS controls the transmit power of femtocell users by
pricing their resulted interference power levels at the MBS receiver
subject to a maximum tolerable interference margin.

\item This paper formulates a \emph{Stackelberg game} to jointly maximize
the revenue of the macrocell and the individual utilities of
different femtocell users for the proposed price-based resource
allocation. More specifically, the interference tolerance margin at
the MBS is used as the resource that the leader (MBS) and the
followers (femtocell users) in the formulated Stackelberg game
compete for, under which simple and effective price-based resource
allocation strategies are obtained. In this paper, we propose two
pricing schemes: \emph{non-uniform pricing} in which different
interference-power prices are assigned to different femtocell users,
and \emph{uniform pricing} in which a uniform price applies to all
the femtocell users. In addition, in the uniform-pricing case, we
develop a \emph{distributed interference bargaining algorithm} that
requires minimal network information exchange between the MBS and
HBSs. We show that the non-uniform pricing scheme is optimal from
the perspective of revenue maximization for the MBS, while the
uniform pricing scheme maximizes the sum-rate of femtocell users.

\item This paper studies the Stackelberg equilibriums for the
proposed power allocation games with non-uniform or uniform pricing
under two types of practical femtocell channel models:
\emph{sparsely deployed scenario} applicable for rural areas in
which the interference channels across different femtocells are
ignored, and \emph{densely deployed scenario} for urban areas in
which the cross-femtocell interference is assumed to be present, but
subject to certain peak power constraint. Moreover, for the sparsely
deployed scenario, we obtain the closed-form expressions for the
optimal interference price and power allocation solutions, while for
the densely deployed scenario, lower and upper bounds on the
achievable revenue for the MBS are obtained by applying the
solutions in the sparsely deployed case.
\end{itemize}

The rest of this paper is organized as follows. Section \ref{System
Model} introduces the system model. Section \ref{Problem
Formulation} formulates the Stackelberg game for price-based
resource allocation. Sections \ref{Sparsely Deployed Scenario} and
\ref{Densely Deployed Scenario} investigate the Stackelberg
equilibriums and the optimal price and power allocation solutions
for the sparsely deployed scenario and densely deployed scenario,
respectively. Section \ref{NumericalResults} provides numerical
examples to validate the proposed studies. Finally, Section
\ref{conclusions} concludes the paper.

\section{System Model}\label{System Model}
In this paper, we consider a two-tier femtocell network consisting
of one central MBS  serving a region $\mathcal {R}$, within which
there are in total $N$ femtocells deployed by home or office users.
It is assumed that all femtocells access the same frequency band as
the macrocell. In each femtocell, there is one dedicated HBS
providing service for several wireless devices. Each wireless device
is regarded as one user in the femtocell network. For analytical
tractability, we assume that at any given frequency band (e.g., one
frequency sub-channel in OFDMA-based femtocells), there is at most
one scheduled active user during each signaling time-slot in each
femtocell, i.e., orthogonal uplink transmission is adopted. In this
paper, we focus our study on the uplink transmission in the
femtocell network over a single frequency band, while it is worth
pointing out that the results obtained under this assumption can be
easily extended to broadband femtocell systems with parallel
frequency sub-channels using the ``dual decomposition'' technique
similarly as \cite{Zhang08MAC}.

\begin{figure}[t]
        \centering
        \includegraphics*[width=12cm]{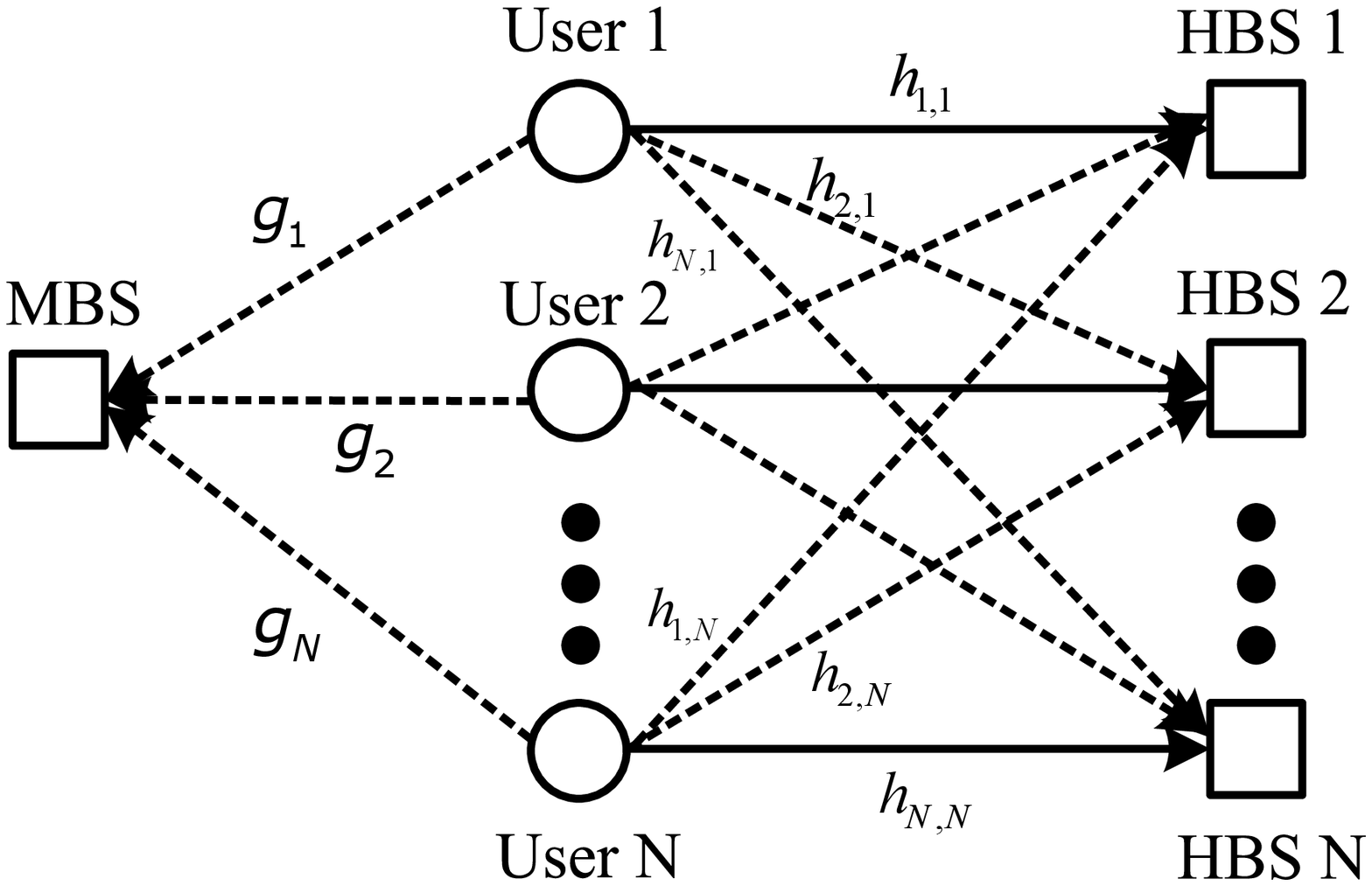}
        \caption{System model of the femtocell network for the uplink transmission.}\vspace{-3mm}
        \label{model}
\end{figure}

Under the above framework, for a given time-slot, the uplink
transmission for the two-tier femtocell network can be described in
Fig. \ref{model}. As shown in Fig. \ref{model}, user $i$ denotes the
scheduled user transmitting to its HBS $\mathcal {B}_i$, where
$i=1,2,\cdots,N$. All the terminals involved are assumed to be
equipped with a single antenna. For the purpose of exposition, all
the channels involved are assumed to be block-fading, i.e., the
channels remain constant during each transmission block, but
possibly change from one block to another. The channel power gain of
the link between user $i$ and HBS $\mathcal {B}_j$ is denoted by
$h_{j,i}$. The channel power gain of the link between user $i$ and
the MBS is given by $g_{i}$. All the channel power gains are assumed
to be independent and identically distributed (i.i.d.) random
variables (RVs) each having a continuous probability density
function (PDF). The additive noises at HBSs and MBS are assumed to
be independent circularly symmetric complex Gaussian (CSCG) RVs,
each of which is assumed to have zero mean and variance $\sigma^2$.

We consider two practical femtocell channel models: \emph{sparsely
deployed scenario} and \emph{densely deployed scenario}. For the
sparsely deployed scenario, we assume that the mutual interference
between the femtocells is neglected. This is because the channel
power gain drops sharply with the increasing of the distance between
femtocells due to path loss (which is proportional to $d^{-\alpha}$,
where $d$ is the distance and $\alpha$ is the path loss exponent).
Besides, since femtocells are usually deployed indoor, the
penetration loss is also significant. Therefore, it is reasonable to
assume that the interference between femtocells can be neglected
when the femtocells are sparsely deployed. In practice, this
scenario is applicable to the femtocell networks deployed in rural
areas where the distances between femtocells are usually large.
While for the urban areas, where the femtocells are close to each
other and thus the mutual interference between femtocells cannot be
ignored, the sparsely deployed scenario may not be suitable. For
such situations, we consider the densely deployed scenario that
takes the mutual interference between different femtocells into
account. Especially, for this scenario, we assume that the aggregate
interference at user $i$'s receiver due to all the other femtocell
users is bounded, i.e., $\sum_{j=1,j\neq i}^{N}I^F_{j}\le
\varepsilon$, where $\varepsilon$ denotes the bound and $I^F_{j}$
denotes the power of the interference from femtocell user $j$. This
assumption is valid due to the following facts: (i) the
cross-femtocell channel power gains are usually very weak due to the
penetration loss; and (ii) the peak transmit power of each femtocell
user is usually limited due to practical constraints on power
amplifiers.

\textbf{Notation:} In this paper, the boldface capital and lowercase
letters are used to denote matrices and vectors, respectively. The
inequalities for vectors are defined element-wise, i.e.,
$\boldsymbol{x}\preceq\boldsymbol{y}$ represents $x_i\le
y_i,~\forall i$, where $x_i$ and $y_i$ are the $i$th elements of the
vector $\boldsymbol{x}$ and $\boldsymbol{y}$, respectively. The
superscript $T$ denotes the transpose operation of a vector.

\section{Problem Formulation} \label{Problem Formulation}
In this section, we first present the Stackelberg game formulation
for the price-based power allocation scheme. Then, the Stackelberg
equilibrium of the proposed game is investigated.

\subsection{Stackelberg Game Formulation}
In this paper, we assume that the maximum interference that the MBS
can tolerate is $Q$, i.e., the aggregate interference from all the
femtocell users should not be larger than $Q$. Mathematically, this
can be written as
\begin{align}
\sum_{i=1}^{N} I_i \le Q,
\end{align}
where $I_i$ denotes the power of the interference from femtocell
user $i$. This constraint is known as \emph{interference power
constraint} or \emph{interference temperature constraint} in CRNs.

Different from the cognitive radio studies, in this paper, we assume
that such an interference power constraint is imposed at the MBS,
which protects itself through pricing the interference from the
femtocell users. The Stackelberg game model \cite{GameTheory1993} is
thus applied in this scenario. Stackelberg game is a strategic game
that consists of a leader and several followers competing with each
other on certain resources. The leader moves first and the followers
move subsequently. In this paper, we formulate the MBS as the
leader, and the femtocell users as the followers. The MBS (leader)
imposes a set of prices on per unit of received interference power
from each femtocell user. Then, the femtocell users (followers)
update their power allocation strategies to maximize their
individual utilities based on the assigned interference prices.

Under the above game model, it is easy to observe that the MBS's
objective is to maximize its revenue obtained from selling the
interference quota to femtocell users. Mathematically, the revenue
of MBS can be calculated by
\begin{align}
U_{MBS}\left(\boldsymbol{\mu},
\boldsymbol{p}\right)=\sum_{i=1}^{N}\mu_i I_i(p_i),
\end{align}
where $\boldsymbol{\mu}$ is the interference price vector with
$\boldsymbol{\mu}=[\mu_1,\mu_2,\cdots,\mu_N]^T$, with $\mu_i$
denoting the interference price for user $i$; $I_i(p_i)$ is the
interference power received from femtocell user $i$, and
$\boldsymbol{p}$ is a vector of power levels for femtocell users
with $\boldsymbol{p}=[p_1,p_2,\cdots,p_N]^T$. Note that $\forall i$,
$p_i$ is actually a function of $\mu_i$ under the Stackelberg game
formulation, which indicates that the amount of the interference
quota that each femtocell user is willing to buy is dependent on its
assigned interference price. Since the maximum aggregate
interference that the MBS can tolerate is limited, the MBS needs to
find the optimal interference prices $\boldsymbol{\mu}$ to maximize
its revenue within its tolerable aggregate interference margin. This
is obtained by solving the following optimization problem:

\underline{\emph{Problem 3.1:}}
\begin{align}
\max_{\boldsymbol{\mu}\succeq \boldsymbol{0}}~&U_{MBS}\left(\boldsymbol{\mu}, \boldsymbol{p}\right),\\
\mbox{s.t.}~~&\sum_{i=1}^{N} I_i(p_i) \le Q.
\end{align}

At the femtocell users' side, the received SINR at HBS $\mathcal
{B}_i$ for user $i$ can be written as
\begin{align}
\gamma_i\left(p_i,\boldsymbol{p}_{-i}\right)=\frac{p_ih_{i,i}}{\sum_{j\neq
i}p_j h_{i,j}+\sigma_i^2}, \forall i \in
\left\{1,2,\cdots,N\right\}. \label{eq-SINR}
\end{align}
where $\sigma_i^2$ is the background noise at HBS $\mathcal {B}_i$
taking into account of the interference from the macrocell users,
and $\boldsymbol{p}_{-i}$ is a vector of power allocation for all
users except user $i$, i.e., $\boldsymbol{p}_{-i}=[p_1,\cdots,
p_{i-1},p_{i+1},\cdots,p_N]^T$. Without loss of generality, it is
assumed for convenience that $\sigma_i^2=\sigma^2, \forall i$ in the
rest of this paper.

The utility for user $i$ can be defined as
\begin{align}
U_i\left(p_i,\boldsymbol{p}_{-i},\mu_i\right)=\lambda_i\log\left(1+\gamma_i\left(p_i,\boldsymbol{p}_{-i}\right)\right)-\mu_i
I_i(p_i),\label{eq-utility-FU}
\end{align}
where $\lambda_i$ is the utility gain per unit transmission rate for
user $i$, and $I_i(p_i)$ is the interference quota that user $i$
intends to buy from the MBS under the interference price $\mu_i$
with $I_i (p_i)\triangleq g_i p_i$. It is observed from
\eqref{eq-utility-FU} that the utility function of each femtocell
user consists of two parts: \emph{profit} and \emph{cost}. If the
femtocell user increases its transmit power, the transmission rate
increases, and so does the profit. On the other hand, with the
increasing of the transmit power, the femtocell user will definitely
cause more interference to the MBS. As a result, it has to buy more
interference quota from the MBS, which increases the cost.
Therefore, power allocation strategies are needed at the femtocell
users to maximize their own utilities. Mathematically, for each user
$i$, this problem can be formulated as

\underline{\emph{Problem 3.2:}}
\begin{align}
\max_{p_i}~&U_i\left(p_i,\boldsymbol{p}_{-i},\boldsymbol{\mu}\right),\\
\mbox{s.t.}~~&~p_i \ge 0.
\end{align}

Problems 3.1 and 3.2 together form a Stackelberg game. The objective
of this game is to find the \textbf{Stackelberg Equilibrium} (SE)
point(s) from which neither the leader (MBS) nor the followers
(femtocell users) have incentives to deviate. The SE for the
proposed game is investigated in the following subsection.

\subsection{Stackelberg Equilibrium}

For the proposed Stackelberg game, the SE is defined as follows.

\underline{\emph{Definition 3.1:}} Let $\boldsymbol{\mu}^*$ be a
solution for Problem 3.1 and $p_i^{*}$ be a solution for Problem 3.2
of the $i$th user. Then, the point
$\left(\boldsymbol{\mu}^*,\boldsymbol{p}^*\right)$ is a SE for the
proposed Stackelberg game if for any
$\left(\boldsymbol{\mu},\boldsymbol{p}\right)$ with
$\boldsymbol{\mu}\succeq \boldsymbol{0}$ and $\boldsymbol{p}\succeq
\boldsymbol{0}$, the following conditions are satisfied:
\begin{align}
U_{MBS}\left(\boldsymbol{\mu}^*,\boldsymbol{p}^*\right)&\ge
U_{MBS}\left(\boldsymbol{\mu},\boldsymbol{p}^*\right),\\U_i\left(p_i^*,\boldsymbol{p}_{-i}^*,\boldsymbol{\mu}^*\right)&\ge
U_i\left(p_i,\boldsymbol{p}_{-i}^*,\boldsymbol{\mu}^*\right),
\forall i.
\end{align}

Generally, the SE for a Stackelberg game can be obtained by finding
its subgame perfect \textbf{Nash Equilibrium} (NE). In the proposed
game, it is not difficult to see that the femtocell users strictly
compete in a noncooperative fashion. Therefore, a noncooperative
power control subgame is formulated at the femtocell users' side.
For a noncooperative game, NE is defined as the operating point(s)
at which no player can improve utility by changing its strategy
unilaterally, assuming everyone else continues to use its current
strategy. At the MBS's side, since there is only one player, the
best response of the MBS can be readily obtained by solving Problem
3.1. To achieve this end, the best response functions for the
followers (femtocell users) must be obtained first, since the leader
(MBS) derives its best response function based on those of the
followers or femtocell users. For the proposed game in this paper,
the SE can be obtained as follows: For a given $\boldsymbol{\mu}$,
Problem 3.2 is solved first. Then, with the obtained best response
functions $\boldsymbol{p}^*$ of the femtocells, we solve Problem 3.1
for the optimal interference price $\boldsymbol{\mu}^*$.

It is not difficult to see that, in the above formulation, we assume
that the MBS charges each femtocell user with a different
interference price. We thus refer to this pricing scheme as
\emph{non-uniform pricing}. In addition, we consider a special case
of this pricing scheme referred to as \emph{uniform pricing}, in
which the MBS charges each femtocell with the same interference
price, i.e., $\mu_i=\mu, \forall i$. In the following, these two
pricing schemes are investigated for the \emph{sparsely deployed
scenario} and the \emph{densely deployed scenario}, respectively.

\section{Sparsely Deployed Scenario} \label{Sparsely Deployed Scenario}
In the sparsely deployed scenario, we assume that the femtocells are
sparsely deployed within the macrocell. Under this assumption, the
mutual interference between any pair of femtocells is negligible and
thus ignored, i.e., $h_{i,j}=0, \forall i\neq j$. In this scenario,
since the inter-femtocell interference is ignored, the problem of
solving price-based resource allocation is simplified, which enables
us to get the closed-form price and power allocation solutions for
the formulated Stackelberg game. As will be shown in the next
section, these solutions will enlighten us on the power allocation
strategies for the more general densely deployed scenario as well.

In this case, SINR given in \eqref{eq-SINR} can be approximated by
\begin{align}
\gamma_i\left(p_i,\boldsymbol{p}_{-i}\right)\approx\frac{p_ih_{i,i}}{\sigma^2},
\forall i \in \left\{1,2,\cdots,N\right\}.
\end{align}

Next, we consider the two pricing schemes: \emph{non-uniform
pricing} and \emph{uniform pricing}, respectively. Then, we compare
these two schemes, highlight their advantages and disadvantages for
implementation, and point out the best situation under which each
scheme should be applied.

\subsection{Non-Uniform Pricing}
For the non-uniform pricing scheme, the MBS sets different
interference prices for different femtocell users. If we denote the
interference price for user $i$ as $\mu_i$, for the sparsely
deployed scenario, Problem 3.2 can be simplified as

\underline{\emph{Problem 4.1:}}
\begin{align}
\max_{p_i}~&\lambda_i\log\left(1+\frac{p_ih_{i,i}}{\sigma^2}\right)-\mu_i
g_ip_i,\\
\mbox{s.t.}~~&~p_i \ge 0.
\end{align}
It is observed that the objective function is a concave function
over $p_i$, and the constraint is affine. Thus, Problem 4.1 is a
convex optimization problem. For a convex optimization problem, the
optimal solution must satisfy the Karush-Kuhn-Tucker (KKT)
conditions. Therefore, by solving the KKT conditions, the optimal
solution for Problem 4.1 can be easily obtained in the following
lemma. Details are omitted for brevity.

\underline{\emph{Lemma 4.1:}} For a given interference price
$\mu_i$, the optimal solution for Problem 4.1 is given by
\begin{align}\label{eq-sparse-op-power}
p_i^*=\left\{\begin{array}{ll}
                       \frac{\lambda_i}{\mu_i
g_i}-\frac{\sigma^2}{h_{i,i}}, &~\mbox{if}~\mu_i<\frac{\lambda_i h_{i,i}}{g_i \sigma^2}, \\
                       0, & ~\mbox{otherwise}.
                     \end{array}
                     \right.
\end{align}
From Lemma 4.1, it is observed that if the interference price is too
high, i.e., $\mu_i>\frac{\lambda_i h_{i,i}}{g_i \sigma^2}$, user $i$
will not transmit. This indicates that user $i$ will be removed from
the game.

We can rewrite the power allocation strategy given in
\eqref{eq-sparse-op-power} as
\begin{align}\label{eq-pi-15}
p_i^*=\left(\frac{\lambda_i}{\mu_i
g_i}-\frac{\sigma^2}{h_{i,i}}\right)^+, ~\forall i,
\end{align}
with $\left(\cdot\right)^+ \triangleq \max\left(\cdot,0\right)$.
Substituting \eqref{eq-pi-15} into Problem 3.1, the optimization
problem at the MBS side can be formulated as

\underline{\emph{Problem 4.2:}}
\begin{align}
\max_{\boldsymbol{\mu}\succcurlyeq
\boldsymbol{0}}~&\sum_{i=1}^{N}\left(\lambda_i-\frac{\mu_i g_i
\sigma^2}{h_{i,i}}\right)^+, \\
\mbox{s.t.}~&\sum_{i=1}^{N}\left(\frac{\lambda_i}{\mu_i}-\frac{g_i
\sigma^2}{h_{i,i}}\right)^+ \le Q.
\end{align}

Note that the above problem is non-convex, since the object function
is a convex function of $\boldsymbol{\mu}$ (maximization of a convex
function is in general non-convex). Nevertheless, it is shown in the
following that this problem can be converted to a series of convex
subproblems.

For user $i$, we introduce the following indicator function
\begin{align}
\chi_i=\left\{\begin{array}{cl}
                1, & \mbox{if}~~\mu_i<\frac{\lambda_i h_{i,i}}{g_i \sigma^2},\\
                0, & \mbox{otherwise}.
              \end{array}
\right.
\end{align}
With the above indicator functions for $i=1,2,\cdots,N$, Problem 4.2
can be reformulated as

\underline{\emph{Problem 4.3:}}
\begin{align}
\max_{\boldsymbol{\chi},~\boldsymbol{\mu}\succcurlyeq
\boldsymbol{0}}~&\sum_{i=1}^{N}\chi_i \left(\lambda_i-\frac{\mu_i
g_i \sigma^2}{h_{i,i}}\right), \\
\mbox{s.t.}~&\sum_{i=1}^{N} \chi_i
\left(\frac{\lambda_i}{\mu_i}-\frac{g_i \sigma^2}{h_{i,i}}\right)
\le Q, \\
&~\chi_i \in\{0,1\}, \forall i,
\end{align}
where $\boldsymbol{\chi}\triangleq[\chi_1, \chi_2, \cdots,
\chi_N]^T$. It is not difficult to see that the above problem is
non-convex due to $\boldsymbol{\chi}$. However, this problem has a
nice property that is explored as follows.  For a given indicator
vector $\boldsymbol{\chi}$, it is easy to verify that Problem 4.3 is
convex.

Next, we consider a special case of Problem 4.3 by assuming that $Q$
is large enough such that all the users are admitted. As a result,
the indicators for all users are equal to 1, i.e.,
$\mu_i<\frac{\lambda_i h_{i,i}}{g_i \sigma^2}, \forall~i.$ Under
this assumption, Problem 4.3 can be transformed to the following
form

\underline{\emph{Problem 4.4:}}
\begin{align}
\min_{\boldsymbol{\mu}\succcurlyeq
\boldsymbol{0}}~&\sum_{i=1}^{N}\frac{\mu_i g_i
\sigma^2}{h_{i,i}}, \\
\mbox{s.t.}~&\sum_{i=1}^{N}\frac{\lambda_i}{\mu_i}\le Q+\sum_{i=1}^N
\frac{g_i \sigma^2}{h_{i,i}} .
\end{align}

Obviously, this problem is convex. The optimal solution of this
problem is given by the following proposition.

\underline{\emph{Proposition 4.1:}} The optimal solution to Problem
4.4 is given by
\begin{align}\label{eq-op-mui}
\mu_i^*=\sqrt{\frac{\lambda_i h_{i,i}}{g_i
\sigma^2}}\frac{\sum_{i=1}^{N}\sqrt{\frac{\lambda_i g_i
\sigma^2}{h_{i,i}}}}{Q+\sum_{i=1}^N \frac{g_i \sigma^2}{h_{i,i}}},
\forall i \in\left\{1,2,\cdots,N\right\}.
\end{align}

\begin{proof} Please refer to Part A of the appendix. \end{proof}

Now, we relate the optimal solution of Problem 4.4 to that of the
original problem, i.e., Problem 4.2, in the following proposition.

\underline{\emph{Proposition 4.2:}} The interference prices given by
\eqref{eq-op-mui} are the optimal solutions of Problem 4.2 \emph{if
and only if} (iff) $Q>\frac{\sum_{i=1}^{N}\sqrt{\frac{\lambda_i g_i
\sigma^2}{h_{i,i}}}}{\min_i \sqrt{\frac{\lambda_i h_{i,i}}{g_i
\sigma^2}}}-\sum_{i=1}^N \frac{g_i \sigma^2}{h_{i,i}}.$

\begin{proof} Please refer to Part B of the appendix. \end{proof}

With the results obtained above, we are now ready for solving
Problem 4.2. The optimal solution of Problem 4.2 is given in the
following theorem.

\underline{\emph{Theorem 4.1:}} Assuming that all the femtocell
users are sorted in the order $\frac{\lambda_1 h_{1,1}}{g_1
\sigma^2}>\cdots>\frac{\lambda_{N-1} h_{{N-1},{N-1}}}{g_{N-1}
\sigma^2}>\frac{\lambda_N h_{N,N}}{g_N \sigma^2}$, the optimal
solution for Problem 4.2 is given by
\begin{align}\label{eq-op-vec-mu}
\boldsymbol{\mu}^*=\left\{\begin{array}{cl}
               \frac{\sum_{i=1}^{N}\sqrt{\frac{\lambda_i g_i
\sigma^2}{h_{i,i}}}}{Q+\sum_{i=1}^N \frac{g_i \sigma^2}{h_{i,i}}}
[\sqrt{\frac{\lambda_1 h_{1,1}}{g_1 \sigma^2}},
\sqrt{\frac{\lambda_2 h_{2,2}}{g_2 \sigma^2}}, \cdots,
\sqrt{\frac{\lambda_N h_{N,N}}{g_N \sigma^2}}]^T, & \mbox{if}\quad
Q>T_N\\
               \frac{\sum_{i=1}^{N-1}\sqrt{\frac{\lambda_i g_i
\sigma^2}{h_{i,i}}}}{Q+\sum_{i=1}^{N-1} \frac{g_i
\sigma^2}{h_{i,i}}} [\sqrt{\frac{\lambda_1 h_{1,1}}{g_1 \sigma^2}},
\cdots, \sqrt{\frac{\lambda_{N-1} h_{N-1,N-1}}{g_{N-1} \sigma^2}},
\infty]^T, &
\mbox{if}\quad T_N\ge Q>T_{N-1}\\
               \vdots & \vdots \\
               \frac{\sum_{i=1}^{2}\sqrt{\frac{\lambda_i g_i
\sigma^2}{h_{i,i}}}}{Q+\sum_{i=1}^{2} \frac{g_i \sigma^2}{h_{i,i}}}
[\sqrt{\frac{\lambda_1 h_{1,1}}{g_1 \sigma^2}},
\sqrt{\frac{\lambda_2 h_{2,2}}{g_2
\sigma^2}},\infty,\cdots,\infty]^T, &
\mbox{if}\quad T_3\ge Q>T_2\\
               \frac{\sqrt{\frac{\lambda_1 g_1
\sigma^2}{h_{1,1}}}}{Q+\frac{g_1 \sigma^2}{h_{1,1}}}
[\sqrt{\frac{\lambda_1 h_{1,1}}{g_1
\sigma^2}},\infty,\cdots,\infty]^T, & \mbox{if}\quad T_2\ge Q>T_1
             \end{array}
\right.,
\end{align}
where $T_N=\frac{\sum_{i=1}^{N}\sqrt{\frac{\lambda_i g_i
\sigma^2}{h_{i,i}}}}{\sqrt{\frac{\lambda_N h_{N,N}}{g_N
\sigma^2}}}-\sum_{i=1}^N \frac{g_i \sigma^2}{h_{i,i}}$,
$T_{N-1}=\frac{\sum_{i=1}^{N-1}\sqrt{\frac{\lambda_i g_i
\sigma^2}{h_{i,i}}}}{\sqrt{\frac{\lambda_{N-1} h_{N-1,N-1}}{g_{N-1}
\sigma^2}}}-\sum_{i=1}^{N-1} \frac{g_i \sigma^2}{h_{i,i}}$,
$\cdots$, $T_2=\frac{\sum_{i=1}^{2}\sqrt{\frac{\lambda_i g_i
\sigma^2}{h_{i,i}}}}{\sqrt{\frac{\lambda_{2} h_{2,2}}{g_{2}
\sigma^2}}}-\sum_{i=1}^{2}\frac{g_i \sigma^2}{h_{i,i}}$, and
$T_1=\frac{\sqrt{\frac{\lambda_1 g_1
\sigma^2}{h_{1,1}}}}{\sqrt{\frac{\lambda_1 h_{1,1}}{g_1
\sigma^2}}}-\frac{g_1 \sigma^2}{h_{1,1}}=0$.

\begin{proof}
If $Q>T_N$, the optimal $\boldsymbol{\mu}^*$ is readily obtained by
Proposition 4.2. For other intervals of $Q$, e.g., $T_{N-1}\le Q\le
T_N$, the proof of the optimality for the corresponding
$\boldsymbol{\mu}^*$ can be obtained similarly as Proposition 4.2,
and is thus omitted. The proof of Theorem 4.1 thus follows.
\end{proof}

\emph{Remark:} From the system design perspective, the results given
in \eqref{eq-op-vec-mu} are very useful in practice. For instance,
if the MBS sets the interference price for a user to $\infty$, this
user will not transmit; however, if the system is designed to admit
all the $N$ femtocell users, the interference tolerance margin $Q$
at the MBS needs to be set to be above $T_N$.

Now, the Stackelberg game for the sparsely deployed scenario with
non-uniform pricing  is completely solved. The SE for this
Stackelberg game is then given as follows.

\underline{\emph{Proposition 4.3:}} The SE for the Stackelberg game
formulated in Problems 4.1 and 4.2 is
$\left(\boldsymbol{\mu}^*,\boldsymbol{p}^*\right)$, where
$\boldsymbol{\mu}^*$ is given by \eqref{eq-op-vec-mu}, and
$\boldsymbol{p}^*$ is given by \eqref{eq-sparse-op-power}.

In practice, the proposed game can be implemented as follows.

\emph{First}, for any femtocell user $i$, the MBS measures its
channel gain, $g_i$, and collects other information such as
$\lambda_i$ and $h_{i,i}$, from HBS $i$ through the backhaul link.
The MBS then computes $\frac{\lambda_i h_{i,i}}{g_i \sigma^2}$ for
all $i$'s and use them to compute the threshold vector
$\boldsymbol{T}=[T_N, T_{N-1}, \cdots, T_1]^T$ by Theorem 4.1.

\emph{Second}, with the obtained threshold vector $\boldsymbol{T}$,
the MBS decides the interference price for each femtocell user based
on its available interference margin $Q$ according to
\eqref{eq-op-vec-mu}. Then, the interference prices are fed back to
femtocell users through the backhaul links between the MBS and the
HBSs.

\emph{Finally}, after receiving the interference prices from their
respective HBSs, the femtocell users decide their transmit power
levels according to \eqref{eq-sparse-op-power}.

Moreover, based on the special structure of \eqref{eq-op-vec-mu}, we
propose the following algorithm to compute the interference prices
for the femtocell users at the MBS.

\underline{\emph{Algorithm 4.1: Successive User Removal}}

\begin{itemize}
\item \emph{Step 1:} Set $K=N$.

\item \emph{Step 2:} Sort the $K$ users according to
$\frac{\lambda_i h_{i,i}}{g_i \sigma^2}$ (i.e., $\frac{\lambda_1
h_{1,1}}{g_1 \sigma^2}>\cdots>\frac{\lambda_{K-1}
h_{{K-1},{K-1}}}{g_{K-1} \sigma^2}>\frac{\lambda_K h_{K,K}}{g_K
\sigma^2}$).

\item \emph{Step 3:} Compute
$q_K=\frac{\sum_{i=1}^{N}\sqrt{\frac{\lambda_i g_i
\sigma^2}{h_{i,i}}}}{Q+\sum_{i=1}^N \frac{g_i \sigma^2}{h_{i,i}}}.$

\item \emph{Step 4:} Comparing the $q_K$ with $\sqrt{\frac{\lambda_K
h_{K,K}}{g_K \sigma^2}}$. If $q_K>\sqrt{\frac{\lambda_K h_{K,K}}{g_K
\sigma^2}}$, remove user $K$ from the game, set $K=K-1$, and go to
Step 3; otherwise, go to Step 5.

\item \emph{Step 5:} With $q_K$ and $K$, the interference price
$\mu_i$ for user $i$ is given by
\begin{align}
\mu_i=\left\{\begin{array}{ll}
               q_K \sqrt{\frac{\lambda_i h_{i,i}}{g_i \sigma^2}}, & \mbox{if}~i\leq K \\
               \infty, & \mbox{otherwise.}
             \end{array}
\right.
\end{align}
\end{itemize}

It is observed from the above algorithm that, to obtain the optimal
interference price vector $\boldsymbol{\mu}^*$, the MBS has to
measure and collect the network state information to compute
$\frac{\lambda_i h_{i,i}}{g_i \sigma^2}$ for each individual
femtocell user $i$. This will incur great implementation complexity
and feedback overhead for the MBS and the HBSs. To relieve this
burden, we must reduce the amount of information that needs to be
known at the MBS. In the following, we consider the uniform pricing
scheme, for which the MBS only needs to measure the total received
interference power $\sum_{i=1}^{N} I_i(p_i)$ from all the femtocell
users to compute the optimal interference price, via a new
\emph{distributed interference price bargaining} algorithm.

\subsection{Uniform Pricing}

For the uniform pricing scheme, the MBS sets a uniform interference
price for all the femtocell users, i.e., $\mu_i=\mu, \forall i$.
With a uniform price $\mu$, the optimal power allocation for
femtocell users can be easily obtained from
\eqref{eq-sparse-op-power} by setting $\mu_i=\mu$, i.e.,
\begin{align}\label{eq-sparse-op-power-uni}
p_i^*=\left(\frac{\lambda_i}{\mu
g_i}-\frac{\sigma^2}{h_{i,i}}\right)^+, ~\forall i.
\end{align}

Then, at the MBS's side, the optimization problem reduces to

\underline{\emph{Problem 4.5:}}
\begin{align}
\max_{\mu>0}~&\sum_{i=1}^{N}\left(\lambda_i-\frac{\mu g_i
\sigma^2}{h_{i,i}}\right)^+, \\
\mbox{s.t.}~&\sum_{i=1}^{N}\left(\frac{\lambda_i}{\mu}-\frac{g_i
\sigma^2}{h_{i,i}}\right)^+ \le Q. \label{eq-IPCon}
\end{align}
This problem has the same structure as Problem 4.2. Therefore, it
can be solved by the same method for Problem 4.2. Details are thus
omitted here for brevity.

\underline{\emph{Corollary 4.1:}} Assuming that all the users are
sorted in the order $\frac{\lambda_1 h_{1,1}}{g_1
\sigma^2}>\cdots>\frac{\lambda_{N-1} h_{{N-1},{N-1}}}{g_{N-1}
\sigma^2}>\frac{\lambda_N h_{N,N}}{g_N \sigma^2}$, the optimal
solution for Problem 4.5 is given by
\begin{align}\label{eq-op-uni-mu}
\mu^*=\left\{\begin{array}{cl}
               \frac{\sum_{i=1}^{N}
\lambda_i}{Q+\sum_{i=1}^{N}\frac{g_i\sigma^2}{h_{i,i}}}, &
\mbox{if}\quad Q>\tilde{T}_N\\
               \frac{\sum_{i=1}^{N-1}
\lambda_i}{Q+\sum_{i=1}^{N-1}\frac{g_i\sigma^2}{h_{i,i}}}, &
\mbox{if}\quad \tilde{T}_N
\ge Q>\tilde{T}_{N-1}\\
               \vdots & \vdots \\
               \frac{
\lambda_1}{Q+\frac{g_1\sigma^2}{h_{1,1}}}, & \mbox{if}\quad
\tilde{T}_2\ge Q>\tilde{T}_1,
             \end{array}
\right.
\end{align}
where $\tilde{T}_N=\frac{\sum_{i=1}^N \lambda_i}{ \frac{\lambda_N
h_{N,N}}{g_N \sigma^2}}-\sum_{i=1}^N \frac{g_i \sigma^2}{h_{i,i}}$,
$\tilde{T}_{N-1}=\frac{\sum_{i=1}^{N-1}
\lambda_i}{\frac{\lambda_{N-1} h_{N-1,N-1}}{g_{N-1}
\sigma^2}}-\sum_{i=1}^{N-1} \frac{g_i \sigma^2}{h_{i,i}}$, $\cdots$,
$\tilde{T}_2=\frac{\sum_{i=1}^{2} \lambda_i}{\frac{\lambda_{2}
h_{2,2}}{g_{2} \sigma^2}}-\sum_{i=1}^{2} \frac{g_i
\sigma^2}{h_{i,i}}$, and
$\tilde{T}_1=\frac{\lambda_1}{\frac{\lambda_1 h_{1,1}}{g_1
\sigma^2}}-\frac{g_1 \sigma^2}{h_{1,1}}=0$.

From Corollary 4.1, it is not to difficult to observe that the
optimal price $\mu^*$ is unique for a given $Q$. Consequently, the
SE for this Stackelberg game is unique and given as follows.

\underline{\emph{Corollary 4.2:}} The SE for the Stackelberg game
for the uniform pricing case is
$\left(\mu^*,\boldsymbol{p}^*\right)$, where $\mu^*$ is given by
\eqref{eq-op-uni-mu}, and $\boldsymbol{p}^*$ is given by
\eqref{eq-sparse-op-power-uni}.

In practice, the Stackelberg game for the uniform-pricing case can
be implemented in the same centralized way as that for the
non-uniform pricing case, which requires the MBS to collect a large
amount of information from each femtocell user. However, Problem 4.5
has some nice properties that can be explored for the algorithm
design. It is observed from Problem 4.5 that both the objective
function and the left hand side of \eqref{eq-IPCon} are
monotonically decreasing functions of $\mu$. Therefore, the
objective function is maximized iff \eqref{eq-IPCon} is satisfied
with equality. By exploiting this fact, we propose the following
algorithm to achieve the SE of the Stackelberg game in the
unform-pricing case.

\underline{\emph{Algorithm 4.2: Distributed Interference Price
Bargaining} }

\begin{itemize}
\item \emph{Step 1:} The MBS initializes the interference price $\mu$,
and broadcasts $\mu$ to all the femtocell users (e.g., through the
HBSs via the backhaul links).

\item \emph{Step 2:} Each femtocell user calculates its optimal
transmit power $p^*_i$ based on the received $\mu$ by
\eqref{eq-sparse-op-power-uni}, and attempts to transmit with
$p^*_i$.

\item \emph{Step 3:} The MBS measures the total received interference
$\sum_{i=1}^{N} I_i(p_i)$, and updates the interference price $\mu$
based on $\sum_{i=1}^{N} I_i(p_i)$. Assume that $\epsilon$ is a
small positive constant that controls the algorithm accuracy. Then,
if $\sum_{i=1}^{N} I_i(p_i)
>Q+\epsilon$, the MBS increases the interference price by $\Delta \mu$;
if $\sum_{i=1}^{N} I_i(p_i) <Q-\epsilon$, the MBS decreases the
interference price by $\Delta \mu$, where $\Delta \mu>0$ is a small
step size. After that, the MBS broadcasts the new interference price
to all the femtocells users.

\item \emph{Step 4:} Step 2 and Step 3 are repeated until
$\big|\sum_{i=1}^{N} I_i(p_i)-Q\big|\le \epsilon$.

\end{itemize}

\emph{Remark:} The convergence of Algorithm 4.2 is guaranteed due to
the following facts: (i) the optimal $\mu$ is obtained when
\eqref{eq-IPCon} is satisfied with equality; and  (ii) the left hand
side of \eqref{eq-IPCon} is a monotonically decreasing function of
$\mu$.

It is seen that Algorithm 4.2 is a distributed algorithm. At the MBS
side, the MBS only needs to measure the total received interference
$\sum_{i=1}^{N} I_i(p_i)$. At the femtocell side, each femtocell
user only needs to know the channel gain to its own HBS to compute
the transmit power. Overall, the amount of information that needs to
be exchanged in the network is greatly reduced, as compared to the
centralized approach.

\subsection{Non-Uniform Pricing vs. Uniform Pricing}

In the following, we summarize the main results on comparing the two
schemes of non-uniform pricing and uniform pricing.

\emph{First, it is observed that the non-uniform pricing scheme must
be implemented in a centralized way, while the uniform pricing
scheme can be implemented in a decentralized way.} Therefore,
uniform pricing is more favorable when the network state information
is not available.

\emph{Secondly, the non-uniform pricing scheme maximizes the revenue
of the MBS, while the uniform pricing scheme maximizes the sum-rate
of the femtocell users.} It is easy to observe that non-uniform
pricing is optimal from the perspective of revenue maximization of
the MBS, as compared to uniform pricing. However, it is not
immediately clear that the uniform pricing scheme is indeed optimal
for the sum-rate maximization of the femtocell users. Hence, the
following proposition affirms this property.

\emph{Proposition 4.4:} For a given interference power constraint
$Q$, the sum-rate of the femtocell users is maximized by the uniform
pricing scheme.

\begin{proof}
Please refer to Part C of the appendix.
\end{proof}

\section{Densely Deployed Scenario}\label{Densely Deployed Scenario}

In this scenario, we assume that the femtocells are densely deployed
within the region covered by the macrocell. Therefore, the mutual
interference between femtocells cannot be neglected. However, as
previously stated in the system model, it is still reasonable to
assume that the aggregate interference at user $i$'s receiver due to
all other femtocell users is bounded, i.e., $\sum_{j\neq i}p_j^*
h_{i,j}\le \varepsilon$, where $\varepsilon$ denotes the upper
bound.

For this scenario, we also consider two pricing schemes:
\emph{non-uniform pricing} and \emph{uniform pricing}, which are
studied in the following two subsections, respectively.

\subsection{Non-Uniform Pricing}
Under the non-uniform pricing scheme, the MBS sets different
interference prices for different femtocell users. If we denote the
interference price for user $i$ as $\mu_i$, the best responses for
the noncooperative game at the femtocell users' side can be obtained
by solving Problem 3.2 as follows.

For given $\boldsymbol{p}_{-i}$ and $\mu_i$, it is easy to verify
that Problem 3.2 is a convex optimization problem. Thus, the best
response function for user $i$ can be obtained by setting
$\frac{\partial
U_i\left(p_i,\boldsymbol{p}_{-i},\mu_i\right)}{\partial p_i}$ to
$0$. Taking the first-order derivative of \eqref{eq-utility-FU}, we
have
\begin{align}\label{eq-utility-FU-deriv}
\frac{\partial
U_i\left(p_i,\boldsymbol{p}_{-i},\mu_i\right)}{\partial
p_i}=\frac{\lambda_i}{\frac{p_i}{\gamma_i\left(p_i,\boldsymbol{p}_{-i}\right)}+p_i}-\mu_i
g_i=0.
\end{align}
Substituting the $\gamma_i\left(p_i,\boldsymbol{p}_{-i}\right)$
given in \eqref{eq-SINR} into \eqref{eq-utility-FU-deriv} yields
\begin{align}\label{eq-dense-op-power}
p_i^*=\left(\frac{\lambda_i}{\mu_i g_i}-\frac{\sum_{j\neq i}p_j^*
h_{i,j}+\sigma^2}{h_{i,i}}\right)^+,\forall i
\in\left\{1,2,\cdots,N\right\}.
\end{align}
For a given interference vector $\boldsymbol{\mu}$,
\eqref{eq-dense-op-power} represents an $N$-user non-cooperative
game. It is easy to verify that, for a given interference vector
$\boldsymbol{\mu}$, there exists at least one NE for the
non-cooperative game defined by \eqref{eq-dense-op-power}. In
general, there are multiple NEs, and thus it is NP-hard to get the
optimal power allocation vector $\boldsymbol{p}^*$.

Fortunately, since the aggregate interference is bounded, we may
consider first the \emph{worst case}, i.e., $\sum_{j\neq i}p_j^*
h_{i,j}=\varepsilon$, $\forall i$. In this case, the best response
functions of all users are decoupled in terms of $p_i$'s. If we
denote $\varepsilon+\sigma^2$ as $\theta$, the revenue maximization
problem at the MBS's side will be exactly the same as Problem 4.2,
with $\sigma^2$ replaced by $\theta$. Therefore, the optimal
interference price vectors can be obtained by Theorem 4.1, with
$\sigma^2$ replaced by $\theta$.

On the other hand, we may also consider the \emph{ideal case}, i.e.,
$\sum_{j\neq i}p_j^* h_{i,j}=0$,  $\forall i$. Then, the revenue
maximization problem at the MBS's side will be exactly the same as
Problem 4.2, and the optimal interference price vector can be
obtained by Theorem 4.1.

It is observed that the method used to solve the sparsely deployed
scenario can be directly applied to solve the densely deployed
scenario by considering the worst case and the ideal case,
respectively. It is not difficult to show that the worst case and
the ideal case serve as the lower bound and the upper bound on the
maximum achievable revenue of the MBS, respectively. Furthermore,
these bounds will get closer to each other with the decreasing of
$\varepsilon$ and eventually collide when $\varepsilon=0$.

\subsection{Uniform Pricing}
Under the uniform pricing scheme with $\mu_i=\mu, \forall i$, the
optimal power allocation for femtocell users can be easily obtained
from \eqref{eq-dense-op-power} as
\begin{align}\label{eq-dense-op-power uniform price}
p_i^*=\left(\frac{\lambda_i}{\mu g_i}-\frac{\sum_{j\neq i}p_j^*
h_{i,j}+\sigma^2}{h_{i,i}}\right)^+,\forall i
\in\left\{1,2,\cdots,N\right\}.
\end{align}
Again, it is NP-hard to get the optimal power allocation vector
$\boldsymbol{p}^*$. Similarly, we can solve this problem by either
considering the worst case or the ideal case, for both of which the
methods used to solve the sparsely deployed scenario can be directly
applied. Details are thus omitted for brevity. Last, it is worth
noting that the distributed interference price bargaining algorithm
(Algorithm 4.2) can also be applied in the case of $\varepsilon>0$;
however, the convergence of this algorithm is no more guaranteed due
to the non-uniqueness of NE solutions for the non-cooperate power
game in (\ref{eq-dense-op-power uniform price}). Nevertheless, the
convergence of this algorithm is usually observed in our numerical
experiments when $\varepsilon$ is sufficiently small.

\section{Numerical Results}\label{NumericalResults}
In this section, several numerical examples are provided to evaluate
the performances of the proposed resource allocation strategies
based on the approach of interference pricing. For simplicity, we
assume that the variance of the noise is 1, and the payoff factors
$\lambda_i, \forall i$ are all equal to 1.

A two-tier spectrum-sharing femtocell network with one MBS and three
femtocells is considered. Without loss of generality, the channel
power gains are chosen as follows: $h_{1,1}=1$, $h_{2,2}=1$,
$h_{3,3}=1$, $g_1=0.01$, $g_2=0.1$, and $g_3=1$. In the following,
the first three examples are for the sparsely deployed scenario,
while the last one is for the densely deployed scenario.

\subsection{Example 1: Uniform Pricing vs. Non-Uniform Pricing: Throughput-Revenue Tradeoff}

\begin{figure}[t]
        \centering
        \includegraphics*[width=12cm]{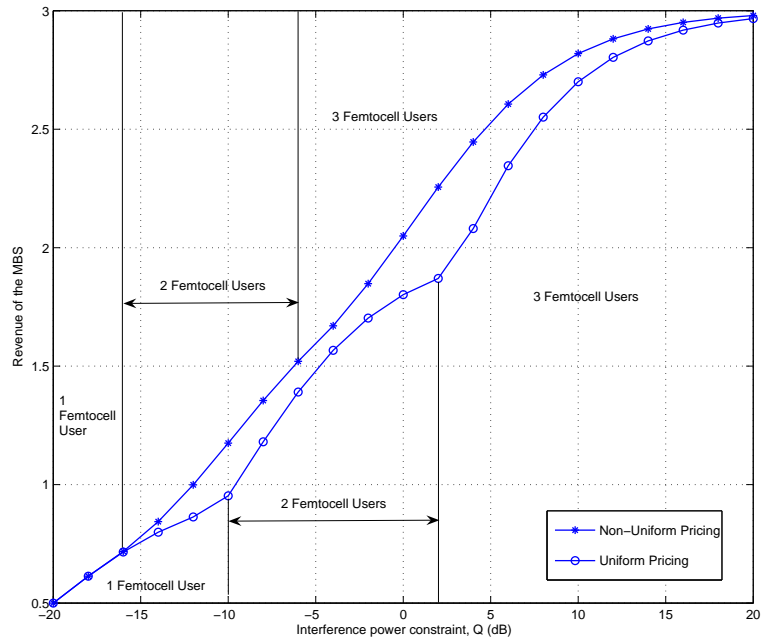}
        \caption{Revenue of the MBS vs. $Q$.}
        \label{Fig1a}
\end{figure}

\begin{figure}[t]
        \centering
        \includegraphics*[width=12cm]{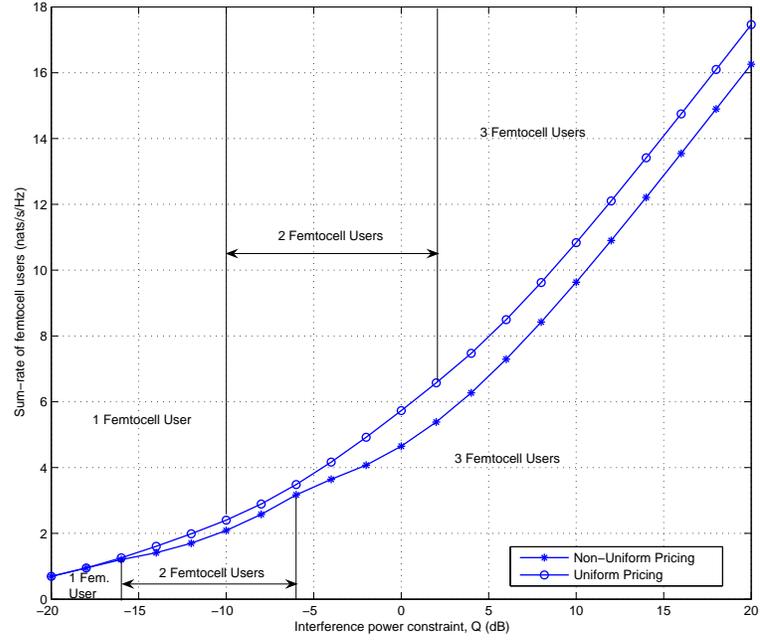}
        \caption{Sum-rate of femtocell users vs. $Q$.}
        \label{Fig1b}
\end{figure}

Figs. \ref{Fig1a} and \ref{Fig1b} show the macrocell revenue and the
sum-rate of femtocell users, respectively, versus the maximum
tolerable interference margin $Q$ at the MBS, with uniform or
non-uniform pricing. It is observed that for the same $Q$, the
revenue of the MBS under the non-uniform pricing scheme is in
general larger than that under the uniform pricing scheme, while the
reverse is generally true for the sum-rate of femtocell users. These
observations are in accordance with our discussions given in Section
\ref{Sparsely Deployed Scenario}. In addition, it is worth noting
that when $Q$ is sufficiently small, the revenues of the MBS become
equal for the two pricing schemes, so are the sum-rates of femtocell
users. This is because when $Q$ is very small, there is only one
femtocell active in the network, and thus by comparing
\eqref{eq-op-vec-mu} and \eqref{eq-op-uni-mu}, the non-uniform
pricing scheme is same as the uniform pricing counterpart in the
single-femtocell case. It is also observed that when $Q$ is
sufficiently large, the revenues of the MBS converge to the same
value for the two pricing schemes. This can be explained as follows.
For the non-uniform pricing scheme, when $Q$ is very large, it is
observed from \eqref{eq-op-vec-mu} that $\mu_i$'s all become very
small, and thus the objective function of Problem 4.2 converges to
$\sum_{i=1}^N \lambda_i$ as $Q\rightarrow \infty$. On the other
hand, for the uniform pricing scheme, the revenue of the MBS can be
written as $\mu^* Q$ at the optimal point, which is equal to
$\frac{Q \sum_{i=1}^{N}
\lambda_i}{Q+\sum_{i=1}^{N}\frac{g_i\sigma^2}{h_{i,i}}}$ when $Q$ is
very large (cf. \eqref{eq-op-uni-mu}). Clearly, this value will
converge to $\sum_{i=1}^N \lambda_i$ as $Q\rightarrow \infty$.

\subsection{Example 2: Comparison of Interference Prices of Femtocell Users under Non-Uniform Pricing}
\begin{figure}[t]
        \centering
        \includegraphics*[width=12cm]{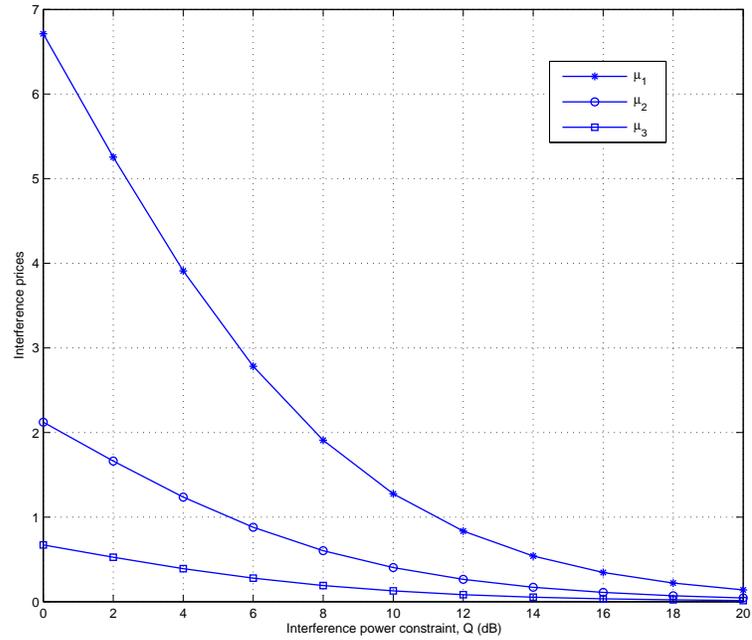}
        \caption{Interference prices for femtocell users vs. $Q$ under non-uniform pricing.}
        \label{Fig1c}
\end{figure}

In this example, we examine the optimal interference prices of the
femtocell users vs. $Q$ under non-uniform pricing. First, it is
observed from Fig. \ref{Fig1c} that, for the same $Q$, the
interference price for femtocell user $1$ is the highest, while that
for femtocell user $3$ is the lowest. This is true due the fact that
$\frac{\lambda_1 h_{1,1}}{g_1 \sigma^2}>\frac{\lambda_2 h_{2,2}}{g_2
\sigma^2}>\frac{\lambda_3 h_{3,3}}{g_3 \sigma^2}$, where a larger
$\frac{\lambda_i h_{i,i}}{g_i \sigma^2}$ indicates that the
corresponding femtocell can achieve a higher profit (transmission
rate) with the same amount network resource (transmit power)
consumed. Therefore, the user with a larger $\frac{\lambda_i
h_{i,i}}{g_i \sigma^2}$ has a willingness to pay a higher price to
consume the network resource. Secondly, it is observed that the
differences between the interference prices decrease with the
increasing of $Q$. This is due to the fact that
$\frac{\sum_{i=1}^{N}\sqrt{\frac{\lambda_i g_i
\sigma^2}{h_{i,i}}}}{Q+\sum_{i=1}^N \frac{g_i \sigma^2}{h_{i,i}}}$
in \eqref{eq-op-vec-mu} decreases with the increasing of $Q$.
Lastly, it is observed that the interference prices for all
femtocell users decrease with the increasing of $Q$, which can be
easily inferred from \eqref{eq-op-vec-mu}. Intuitively, this can be
explained by the practical rule of thumb that a seller would like to
price lower if it has a large amount of goods to sell.

\subsection{Example 3: Convergence Performance of Distributed Interference Price Bargaining Algorithm}
\begin{figure}[t]
        \centering
        \includegraphics*[width=12cm]{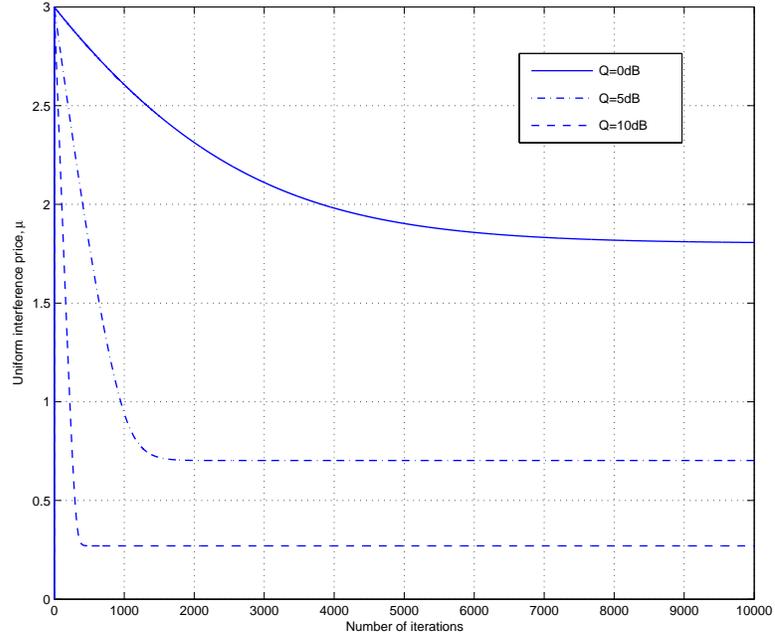}
        \caption{Convergence performance of the distributed interference price bargaining algorithm with the subgradient search.}
        \label{Fig3a}
\end{figure}

In this example, we investigate the convergence performance of the
distributed interference price bargaining algorithm (Algorithm 4.2).
The initial value of $\mu$ is chosen to be $0.001$. The $\Delta \mu$
is chosen to be $0.001\times|\sum_{i=1}^{N} I_i(p_i)-Q|$. The
desired accuracy $\epsilon$ is chosen to be $10^{-6}$. It is
observed from Fig. \ref{Fig3a} that the distributed bargaining
algorithm converges for all values of $Q$. It is also observed that
the convergence speed increases with the increasing of $Q$. This is
because $\Delta \mu$ is proportional to $|\sum_{i=1}^{N}
I_i(p_i)-Q|$, i.e., increasing $Q$ is equivalent to increasing the
step size $\Delta \mu$, and consequently increases the convergence
speed.

\begin{figure}[t]
        \centering
        \includegraphics*[width=12cm]{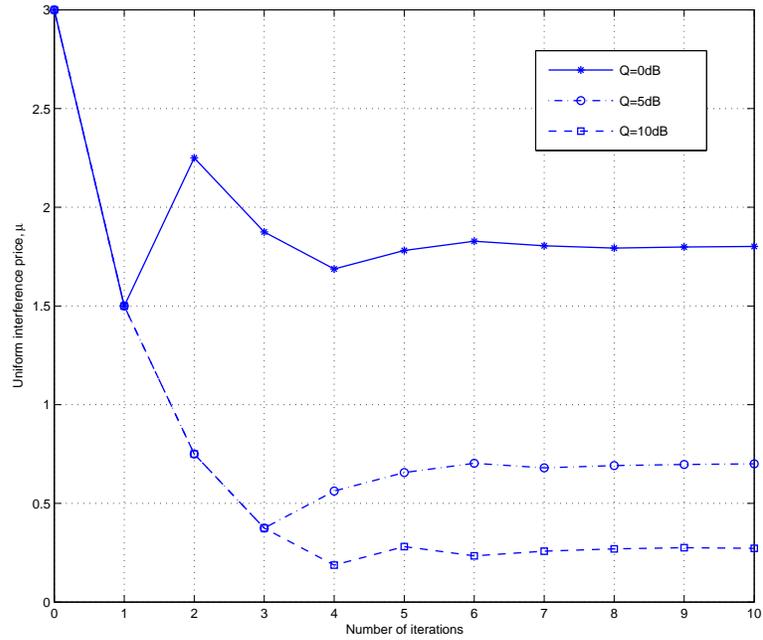}
        \caption{Convergence performance of the distributed interference price bargaining algorithm with the bisection search.}
        \label{Fig3b}
\end{figure}

Actually, the convergence speed of the distributed bargaining
algorithm can be greatly improved by implementing it by the
bisection method, for which the implementation procedure is as
follows. First, the MBS initializes a lower bound $\mu_L$ and an
upper bound $\mu_H$ of the interference price. Then, the MBS
computes $\mu_M=(\mu_L+\mu_H)/2$ and  broadcasts $\mu_M$ to
femtocell users. Receiving $\mu_M$, femtocell users compute their
optimal transmit power and then transmit with the computed power.
The MBS then measures the total received interference
$\sum_{i=1}^{N} I_i(p_i)$ from femtocell users. If $\sum_{i=1}^{N}
I_i(p_i)<Q$, the MBS sets $\mu_H=\mu_M$; otherwise, the MBS sets
$\mu_L=\mu_M$. Then, $ \mu_M$ is recomputed based on the new lower
and upper bounds. The algorithm stops when $|\sum_{i=1}^{N}
I_i(p_i)-Q|$ is within the desired accuracy. It is observed from
Fig. \ref{Fig3b} that the bisection method converges much faster
than the simple subgradient-based method in Fig. \ref{Fig3a}.

\subsection{Example 4: Densely Deployed Scenario under Unform Pricing}
\begin{figure}[t]
        \centering
        \includegraphics*[width=12cm]{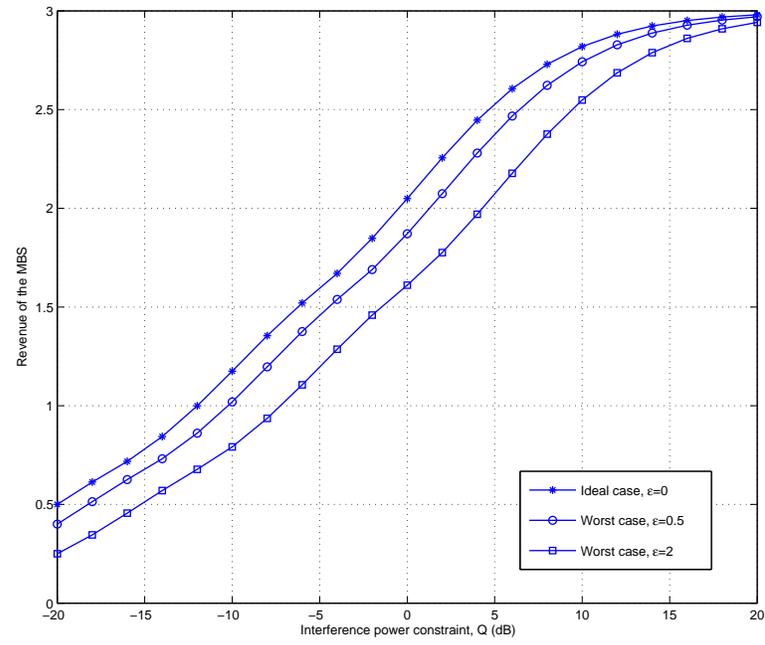}
        \caption{Revenue of the MBS vs. $Q$ for the densely deployed scenario.}
        \label{Fig2}
\end{figure}

In this example, we investigate the macrocell revenue for the
densely deployed scenario under uniform pricing. First, it is
observed from Fig. \ref{Fig2} that the ideal case of $\varepsilon=0$
has the largest revenue of the MBS, compared to the other two cases
with $\varepsilon=0.5, 2$. This verifies that the ideal case can
serve as a revenue upper bound for the densely deployed scenario.
Secondly, the revenues of the MBS for all the three cases of
$\varepsilon=0, 0.5, 2$ increase with the increasing of $Q$,
similarly as expected for the sparsely deployed scenario. Lastly,
the revenue of the MBS is observed to increase with the decreasing
of $\varepsilon$ for the same $Q$, and the revenue differences
become smaller as $Q$ increases.

\section{Conclusion}\label{conclusions}

In this paper, price-based power allocation strategies are
investigated for the uplink transmission in a spectrum-sharing-based
two-tier femtocell network using game theory. An interference power
constraint is applied to guarantee the quality-of-service (QoS) of
the MBS. Then, the Stackelberg game model is adopted to jointly
study the utility maximization of the MBS and femtocell users. The
optimal resource allocation schemes including the optimal
interference prices and the optimal power allocation strategies are
examined. Especially, closed-form solutions are obtained for the
sparsely deployed scenario. Besides, a distributed algorithm that
rapidly converges to the Stackelberg equilibrium is proposed for the
uniform pricing scheme. It is shown that the proposed algorithm has
a low complexity and requires minimum information exchange between
the MBS and femtocell users. The results of this paper will be
useful to the practical design of interference control in
spectrum-sharing femtocell networks.

\section*{Appendix}

\subsection{Proof of Proposition 4.1} \label{App-P41}
It is easy to observe that Problem 4.4 is a convex optimization
problem. Thus, the dual gap between this problem and its dual
optimization problem is zero. Therefore, we can solve Problem 4.4 by
solving its dual problem.

The Lagrangian associated with Problem 4.4 can be written as
\begin{align}
\mathcal
{L}\left(\boldsymbol{\mu},\alpha,\boldsymbol{\beta}\right)=\sum_{i=1}^{N}\frac{\mu_i
g_i \sigma^2}{h_{i,i}}+\alpha
\left(\sum_{i=1}^{N}\frac{\lambda_i}{\mu_i}-Q-\sum_{i=1}^N \frac{g_i
\sigma^2}{h_{i,i}}\right)-\sum_{i=1}^N\beta_i \mu_i,
\end{align}
where $\alpha$ and $\beta_i$ are non-negative dual variables
associated with the constraints
$\sum_{i=1}^{N}\frac{\lambda_i}{\mu_i}\le Q+\sum_{i=1}^N \frac{g_i
\sigma^2}{h_{i,i}}$ and $\mu_i\ge 0$, respectively.

The dual function is then defined as $ \mathcal
{G}\left(\boldsymbol{\mu},\alpha,\boldsymbol{\beta}\right)=
\max_{\boldsymbol{\mu}\succcurlyeq \boldsymbol{0}}
{L}\left(\boldsymbol{\mu},\alpha,\boldsymbol{\beta}\right), $ and
the dual problem is given by $ \min_{\alpha \ge 0,
\boldsymbol{\beta}\succcurlyeq \boldsymbol{0}}\mathcal
{G}\left(\boldsymbol{\mu},\alpha,\boldsymbol{\beta}\right). $ Then,
the KKT conditions can be written as follows:
\begin{align}
\frac{\partial
{L}\left(\boldsymbol{\mu},\alpha,\boldsymbol{\beta}\right)}{\partial
\mu_i}&=0, \forall i, \label{eq-partial-mui}\\
\alpha \left(\sum_{i=1}^{N}\frac{\lambda_i}{\mu_i}-Q-\sum_{i=1}^N
\frac{g_i \sigma^2}{h_{i,i}}\right)&=0, \label{eq-KKT-eq1}\\
\beta_i \mu_i&=0, \forall i,\label{eq-KKT-eq2}\\
\alpha&\ge 0, \\
\beta_i&\ge 0, \forall i,\\
\mu_i&\ge 0, \forall i,\\
\sum_{i=1}^{N}\frac{\lambda_i}{\mu_i}-Q-\sum_{i=1}^N \frac{g_i
\sigma^2}{h_{i,i}}&\le 0 \label{eq-KKT-neq}.
\end{align}

From \eqref{eq-partial-mui}, we have
\begin{align}
\frac{\partial
{L}\left(\boldsymbol{\mu},\alpha,\boldsymbol{\beta}\right)}{\partial
\mu_i}=\frac{g_i \sigma^2}{h_{i,i}}-\alpha
\frac{\lambda_i}{\mu_i^2}-\beta_i, \forall i.
\end{align}
Setting the above function equal to $0$ yields
\begin{align}\label{eq-mui}
\mu_i^2=\alpha \frac{\lambda_i}{\frac{g_i
\sigma^2}{h_{i,i}}-\beta_i},\forall i.
\end{align}

\underline{\emph{Lemma 1:}} $\beta_i=0, \forall i.$

\begin{proof}
Suppose that $\beta_i\neq 0$ for any arbitrary $i$. Then, according
to \eqref{eq-KKT-eq2}, it follows that $\mu_i=0$. From
\eqref{eq-mui}, we know that $\mu_i=0$ indicates that $\alpha=0$,
since $\lambda_i>0$. Then, from \eqref{eq-mui}, it follows that
$\mu_i=0, \forall i$, which contradicts \eqref{eq-KKT-neq}.
Therefore, the preassumption that $\beta_i\neq 0$ for any given $i$
does not hold, and we thus have $\beta_i=0, \forall i.$
\end{proof}

\underline{\emph{Lemma 2:}}
$\sum_{i=1}^{N}\frac{\lambda_i}{\mu_i}-Q-\sum_{i=1}^N \frac{g_i
\sigma^2}{h_{i,i}}=0.$

\begin{proof}
Suppose that $\sum_{i=1}^{N}\frac{\lambda_i}{\mu_i}-Q-\sum_{i=1}^N
\frac{g_i \sigma^2}{h_{i,i}}\neq 0.$ Then, from \eqref{eq-KKT-eq1},
we have $\alpha=0$. Then, from \eqref{eq-mui}, it follows $\mu_i=0,
\forall i$, which contradicts \eqref{eq-KKT-neq}. Therefore, the
aforementioned preassumption does not hold, and we have
$\sum_{i=1}^{N}\frac{\lambda_i}{\mu_i}-Q-\sum_{i=1}^N \frac{g_i
\sigma^2}{h_{i,i}}=0.$
\end{proof}

According to Lemma 1 and $\mu_i\ge 0$, \eqref{eq-mui} can be
rewritten as
\begin{align}
\mu_i=\sqrt{\alpha \frac{\lambda_i h_{i,i}}{g_i \sigma^2}},\forall
i.
\end{align}

Substituting the above equation into (41) and according to Lemma 2,
we have
\begin{align}
\sqrt{\alpha}=\frac{\sum_{i=1}^{N}\sqrt{\frac{\lambda_i g_i
\sigma^2}{h_{i,i}}}}{Q+\sum_{i=1}^N \frac{g_i \sigma^2}{h_{i,i}}}.
\end{align}

Then, substituting (45) back to (44) yields
\begin{align}
\mu_i=\sqrt{\frac{\lambda_i h_{i,i}}{g_i
\sigma^2}}\frac{\sum_{i=1}^{N}\sqrt{\frac{\lambda_i g_i
\sigma^2}{h_{i,i}}}}{Q+\sum_{i=1}^N \frac{g_i \sigma^2}{h_{i,i}}}.
\end{align}

Proposition 4.1 is thus proved.

\subsection{Proof of Proposition 4.2}\label{App-P42}

First, consider the proof of the ``if'' part. It is observed that
the interference vector $\boldsymbol{\mu}^*$ given by
\eqref{eq-op-mui} is the optimal solution of Problem 4.2 if all the
indicator functions are equal to 1, i.e., $\mu_i<\frac{\lambda_i
h_{i,i}}{g_i \sigma^2}, \forall~i \in\left\{1,2,\cdots,N\right\}$.

Substituting \eqref{eq-op-mui} into the above inequalities yields
\begin{align}
\sqrt{\frac{\lambda_i h_{i,i}}{g_i
\sigma^2}}\frac{\sum_{i=1}^{N}\sqrt{\frac{\lambda_i g_i
\sigma^2}{h_{i,i}}}}{Q+\sum_{i=1}^N \frac{g_i
\sigma^2}{h_{i,i}}}<\frac{\lambda_i h_{i,i}}{g_i \sigma^2},
\forall~i \in\left\{1,2,\cdots,N\right\}.
\end{align}
Then, it follows
\begin{align}\label{eq-62}
Q>\frac{\sum_{i=1}^{N}\sqrt{\frac{\lambda_i g_i
\sigma^2}{h_{i,i}}}}{\sqrt{\frac{\lambda_i h_{i,i}}{g_i
\sigma^2}}}-\sum_{i=1}^N \frac{g_i \sigma^2}{h_{i,i}}, \forall~i
\in\left\{1,2,\cdots,N\right\}.
\end{align}
Furthermore, the inequalities given in \eqref{eq-62} can be
compactly written as
\begin{align}
Q>\frac{\sum_{i=1}^{N}\sqrt{\frac{\lambda_i g_i
\sigma^2}{h_{i,i}}}}{\min_i \sqrt{\frac{\lambda_i h_{i,i}}{g_i
\sigma^2}}}-\sum_{i=1}^N \frac{g_i \sigma^2}{h_{i,i}}.
\end{align}
The ``if'' part is thus proved.

Next, consider the ``only if'' part, which is proved by
contradiction as follows.

For the ease of exposition, we assume that femtocell users are
sorted by the following order:
\begin{align}\frac{\lambda_1 h_{1,1}}{g_1
\sigma^2}>\cdots>\frac{\lambda_{N-1} h_{{N-1},{N-1}}}{g_{N-1}
\sigma^2}>\frac{\lambda_N h_{N,N}}{g_N \sigma^2}.\end{align} Then,
in Proposition 4.2, the condition becomes \begin{align}Q>T_N,~
\mbox{where}~ T_N=\frac{\sum_{i=1}^{N}\sqrt{\frac{\lambda_i g_i
\sigma^2}{h_{i,i}}}}{\sqrt{\frac{\lambda_N h_{N,N}}{g_N
\sigma^2}}}-\sum_{i=1}^N \frac{g_i \sigma^2}{h_{i,i}}.\end{align}

Now, suppose $T_{N-1}< Q \le T_{N}$, where $T_{N-1}$ is a threshold
shown later in \eqref{eq-T-N-1}. Suppose that $\boldsymbol{\mu}^*$
given by (\ref{eq-op-mui}) is still optimal for Problem 4.2 with
$T_{N-1}< Q \le T_{N}$. Then, since $Q \le T_{N}$, from
(\ref{eq-op-mui}) it follows that $\mu^*_N\ge\frac{\lambda_N
h_{N,N}}{g_N \sigma^2}$ and thus
$\left(\frac{\lambda_N}{\mu_N^*}-\frac{g_N
\sigma^2}{h_{N,N}}\right)^+=0$. From Problem 4.2, it then follows
that $\mu^*_1,\ldots,\mu^*_{N-1}$ must be the optimal solution of
the following problem
\begin{align}
\max_{\boldsymbol{\mu}\succcurlyeq
\boldsymbol{0}}~&\sum_{i=1}^{N-1}\left(\lambda_i-\frac{\mu_i g_i
\sigma^2}{h_{i,i}}\right)^+, \\
\mbox{s.t.}~&\sum_{i=1}^{N-1}\left(\frac{\lambda_i}{\mu_i}-\frac{g_i
\sigma^2}{h_{i,i}}\right)^+\le Q.
\end{align}

This problem has the same structure as Problem 4.2. Thus, from the
proof of the previous ``if'' part, we can show that the optimal
solution for this problem is given by
\begin{align}\label{eq-op-mu-N-1}
\mu_i^{\star}=\sqrt{\frac{\lambda_i h_{i,i}}{g_i
\sigma^2}}\frac{\sum_{i=1}^{N-1}\sqrt{\frac{\lambda_i g_i
\sigma^2}{h_{i,i}}}}{Q+\sum_{i=1}^{N-1} \frac{g_i
\sigma^2}{h_{i,i}}}, \quad\forall i \in
\left\{1,2,\cdots,N-1\right\},
\end{align}
if $Q>T_{N-1}$, where $T_{N-1}$ is obtained as the threshold for $Q$
above which $\mu^{\star}_i<\frac{\lambda_i h_{i,i}}{g_i \sigma^2}$
holds $\forall i\in\{1,\ldots,N-1\}$, i.e.,
\begin{align}\label{eq-T-N-1}
T_{N-1}=\frac{\sum_{i=1}^{N-1}\sqrt{\frac{\lambda_i g_i
\sigma^2}{h_{i,i}}}}{\sqrt{\frac{\lambda_{N-1} h_{N-1,N-1}}{g_{N-1}
\sigma^2}}}-\sum_{i=1}^{N-1} \frac{g_i \sigma^2}{h_{i,i}}.
\end{align}

Obviously, the optimal interference price solution in
\eqref{eq-op-mu-N-1} for the above problem is different from
$\boldsymbol{\mu}^*$ given by \eqref{eq-op-mui}. Thus, this
contradicts with our presumption that $\boldsymbol{\mu}^*$ is
optimal for Problem 4.2 with $T_{N-1}< Q \le T_{N}$. Therefore, the
interference vector $\boldsymbol{\mu}^*$ given by \eqref{eq-op-mui}
is the optimal solution of Problem 4.2 only if $Q>T_{N}$. The ``only
if'' part thus follows.

By combining the proofs of both the ``if'' and ``only if'' parts,
Proposition 4.2 is thus proved.

\subsection{Proof of Proposition 4.3} \label{App-P43}
For a given interference power constraint $Q$, the sum-rate
maximization problem of the femtocell network can be formulated as
\begin{align}
\max_{\boldsymbol{p}\succcurlyeq
\boldsymbol{0}}~&\sum_{i=1}^{N} \log\left(1+\frac{h_{i,i}p_i}{\sigma_i^2}\right), \\
\mbox{s.t.}~&\sum_{i=1}^{N}g_i p_i \le Q.
\end{align}
It is easy to observe that the sum-rate optimization problem is a
convex optimization problem. The Lagrangian associated with this
problem can be written as
\begin{align}
\mathcal {L}\left(\boldsymbol{p},\nu\right)=\sum_{i=1}^{N}
\log\left(1+\frac{h_{i,i}p_i}{\sigma_i^2}\right)-\nu\left(\sum_{i=1}^{N}g_i
p_i-Q\right),
\end{align}
where $\nu$ is the non-negative dual variable associated with the
constraint $\sum_{i=1}^{N}g_i p_i \le Q$.

The dual function is then defined as $ \mathcal
{G}\left(\boldsymbol{p},\nu\right)= \max_{\boldsymbol{p}\succcurlyeq
\boldsymbol{0}} \mathcal {L}\left(\boldsymbol{p},\nu\right), $ and
the dual problem is $ \min_{\nu\ge 0}\mathcal
{G}\left(\boldsymbol{p},\nu\right). $ For a fixed $\nu$, it is not
difficult to observe that the dual function can also be written as
\begin{align}
\mathcal
{G}\left(\boldsymbol{p},\nu\right)=\max_{\boldsymbol{p}\succcurlyeq
\boldsymbol{0}}\sum_{i=1}^{N} \tilde{\mathcal
{L}}\left(p_i,\nu\right)+\nu Q,
\end{align}
where
\begin{align}
\tilde{\mathcal
{L}}\left(p_i,\nu\right)=\log\left(1+\frac{h_{i,i}p_i}{\sigma_i^2}\right)-\nu
g_i p_i.
\end{align}

Thus, the dual function can be obtained by solving a set of
independent sub-dual-functions each for one user. This is also known
as the ``dual decomposition'' \cite{Zhang08MAC}. For a particular
user, the problem can be expressed as
\begin{align}
\max_{p_i>0}~\log\left(1+\frac{h_{i,i}p_i}{\sigma_i^2}\right)-\nu
g_i p_i.
\end{align}

It can be seen that the dual variable $\nu$ plays the same role as
the uniform price $\mu$. It is easy to observe that these
sub-problems are exactly the same as the power allocation problems
under the uniform pricing scheme when $\nu=\mu$. Note that for the
sum-rate maximization problem, $\nu$ is obtained when the
interference constraint is met with equality. Therefore, the optimal
dual solution of $\nu$ is guaranteed to converge to $\mu^*$ for the
formulated Stackelberg game with uniform pricing.

Proposition 4.3 is thus proved.


\newpage

\end{document}